\documentclass[sigconf]{acmart}
\AtBeginDocument{%
  }

\usepackage{tcolorbox}
\usepackage{colortbl}
\usepackage{multirow}
\usepackage{makecell}
\usepackage{booktabs}

\definecolor{c1}{RGB}{192, 0, 0}
\definecolor{c2}{RGB}{0, 112, 192}
\definecolor{c3}{RGB}{0, 0, 0}

\newcommand{\first}[1]{\textbf{#1}}
\newcommand{\second}[1]{\underline{#1}}

\setcopyright{none}
\renewcommand{\footnotetextcopyrightpermission}[1]{}
\acmConference[Conference acronym 'XX]{Make sure to enter the correct
  conference title from your rights confirmation email}{June 03--05,
  2018}{Woodstock, NY}
\acmISBN{978-1-4503-XXXX-X/2018/06}

\begin{document}

\title[CCFormer: Efficient Cross-Field Interaction and Sequence Compression]{CCFormer: Efficient Cross-Field Interaction and Hierarchical Sequence Compression for Industrial Recommendation}

\author{Yunlong Wang}
\authornote{Equal contribution.}
\email{aloyswang@tencent.com}
\affiliation{%
  \institution{Platform and Content Group, Tencent}
  \country{}
}

\author{Huizhe Zhang}
\email{huizhezhang@tencent.com}
\authornotemark[1]
\affiliation{%
  \institution{Platform and Content Group, Tencent}
  \country{}
}

\author{Haonan Hu}
\email{haonanhu@tencent.com}
\authornotemark[1]
\affiliation{%
  \institution{Platform and Content Group, Tencent}
  \country{}
}

\author{Yudong Li}
\email{elsonli@tencent.com}
\authornote{Corresponding author.}
\affiliation{%
  \institution{Platform and Content Group, Tencent}
  \country{}
}

\author{Bing Wen}
\affiliation{%
  \institution{Platform and Content Group, Tencent}
  \country{}
}

\author{Jianchao Tu}
\affiliation{%
  \institution{Platform and Content Group, Tencent}
  \country{}
}

\author{Chengxiang Zhuo}
\affiliation{%
  \institution{Platform and Content Group, Tencent}
  \country{}
}

\author{Zang Li}
\affiliation{%
  \institution{Platform and Content Group, Tencent}
  \country{}
}


\begin{abstract}
Industrial candidate-ranking systems must combine long user behavior histories with heterogeneous user and item features under strict computational budgets. We propose CCFormer, a structured interaction architecture that separates cross-field retrieval from intra-history modeling. Feature-field-separated cross-attention allows user and target queries to read an evolving history hierarchy, while local subspace token mixing contextualizes history tokens independently of candidate targets. Hierarchical sequence compression reduces the number of history tokens and provides multi-scale historical context for subsequent interactions. Experiments on two public benchmarks show higher mean AUC than the evaluated baselines, while experiments on a large-scale industrial dataset show a favorable accuracy-compute trade-off and higher training throughput than HSTU. Online A/B tests across four recommendation and advertising scenarios report system-level improvements of up to 3.57\% in click-through rate, 3.86\% in page views, and 1.71\% in advertising revenue relative to the corresponding production baselines. CCFormer has been fully deployed in all four scenarios, demonstrating its practical scalability and effectiveness in large-scale recommendation systems.
\end{abstract}



\maketitle

\section{Introduction}
\label{section-1}

Industrial candidate ranking predicts user responses to candidate items using behavior histories and heterogeneous user and item features, with applications in online advertising, e-commerce, news feeds, and short-video platforms \cite{zhou2018deep, yang2022click}. In this work, we study history-conditioned candidate ranking rather than next-item retrieval over an entire item catalog. Recent studies increasingly exploit longer behavior histories to represent more diverse and evolving user interests \cite{kang2018self, li2021lightweight, tan2021sparse}.

\begin{figure}[t]
\centering
  \includegraphics[width=\linewidth]{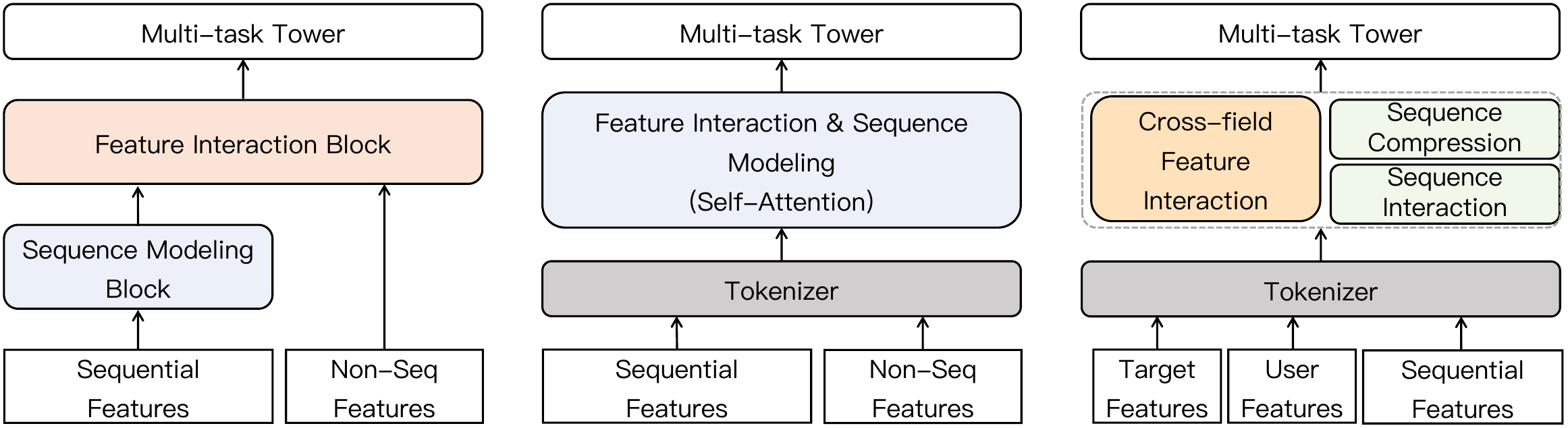}
\caption{Architectural comparison.}
\label{fig:1}
\end{figure}

Due to the strong expressive power over user behavior sequences, Transformer-based models have become a dominant architecture \cite{chai2025longer, dai2025onepiece, tang2026think}. This line of models built upon self-attention can adaptively identify informative historical interactions and capture high-order interactions between past behaviors and the target item. 
Moreover, recent studies suggest that recommendation models can also benefit from scaling up, including increasing the model size, extending the sequence length and allocating more computational budget \cite{zhai2024actions, zhang2026onetrans}. Such scaling trends make it possible to exploit long-term user histories more thoroughly, thereby improving the expressiveness of sequential user representations. 
However, extending Transformer-based sequential models to longer histories can substantially increase training and serving costs under industrial resource constraints. Dense self-attention requires quadratic pairwise computation with respect to sequence length, while its practical memory and latency costs also depend on the attention implementation \cite{gu2026deep}. This challenge is especially critical in industrial ranking scenarios, where models must process massive traffic under strict latency and resource constraints. Consequently, directly extending the input sequence length or increasing model capacity often incurs unacceptable training and inference costs, making Transformer-based sequential models difficult to scale in practical systems.

Representative strategies for reducing the cost of long-sequence recommendation include sequence compression, selective retrieval or truncation, and sparse interaction patterns \cite{yang2026sparse, ma2026blossomrec, lai2026unleashing}. 
As shown in Figure~\ref{fig:1}, some conventional approaches compress long behavior sequences into compact sequence representations before feature interaction. Although such sequence compression reduces the computational cost, it may lose fine-grained behavioral signals and weaken the subsequent interaction between historical behaviors and target items. 
In recent years, some emerging methods retrieve or truncate a target-relevant subsequence and preserve token-level modeling within the limited window~\cite{pi2020search, chang2023twin}, or adopt sparse attention to reduce the cost of long-sequence modeling~\cite{yuan2025native, xia2025training}.
Sequence truncation can exclude useful historical evidence, while dense self-attention within the retained sequence can still incur substantial computational overhead.
In addition, full-sequence modeling involves \emph{cross-field interaction} across semantic fields and \emph{intra-sequence interaction} among historical behaviors. For a target query attending jointly to user and history tokens, a shared softmax normalizes both fields within one distribution. Separately normalizing the two relations offers an alternative inductive bias that decouples their attention normalization before evidence fusion. Meanwhile, exhaustive interactions among history tokens remain costly for long sequences \cite{chen2025pinfm, liu2026est}. These observations motivate an architectural question: \textit{how should cross-field and intra-sequence interactions be organized to capture their distinct roles while efficiently modeling long behavior histories?}

To address these challenges, we propose \textbf{CCFormer}, a structured interaction architecture for industrial candidate ranking with long behavior histories. Feature-field-separated cross-attention enables user and target tokens to retrieve evidence from the retained input history at the first block and from its evolving representations at subsequent blocks. Long-sequence subspace token mixing constructs candidate-independent contextualized history representations, from which target queries retrieve candidate-specific evidence through cross-attention. Hierarchical sequence compression progressively reduces the number of history tokens and broadens the historical spans represented by deeper tokens, lowering the cost of subsequent interactions while providing coarser historical context.
Our contributions are summarized as follows:
\begin{itemize}
    \item We propose a structured interaction architecture for industrial candidate ranking that separates cross-field retrieval from intra-history modeling. Three directed attention pathways---user-to-history, target-to-history, and target-to-user---use relation-specific query projections and separately normalized attention maps, followed by output fusion.
    \item We combine local subspace token mixing over the retained input history with hierarchical sequence compression. Successive compression stages allow deeper mixers to operate on coarser history representations with fewer tokens and broader historical context, while user and target queries read the evolving hierarchy across blocks.
    \item We evaluate CCFormer on two public benchmarks and a large-scale industrial dataset, examining predictive performance and accuracy--compute trade-offs over the tested configurations. Online A/B tests in four production scenarios provide evidence for the business and serving-throughput benefits of the deployed systems relative to their respective baselines.
\end{itemize}

\section{Related Work}

\paragraph{Conventional Deep Learning Recommendation Models.} 
Conventional deep learning recommendation models (DLRMs) strive to learn effective representations from high-dimensional sparse categorical features and dense numerical features \cite{wang2017deep, lian2018xdeepfm, feng2019deep, chen2019behavior}. A common paradigm is to project sparse features into low-dimensional embedding spaces, and then combine them with dense features for prediction. These models design different feature interaction mechanisms to improve the expressiveness of recommendation models. 
For example, Wide \& Deep \cite{cheng2016wide} jointly memorizes low-order feature co-occurrence patterns and generalizes through deep neural networks. DeepFM \cite{guo2017deepfm} integrates factorization machines with deep networks to capture both low-order and high-order feature interactions. 
However, conventional DLRMs usually treat user historical behaviors as aggregated or short-range features, rather than explicitly modeling long-term sequential dependencies. Simply relying on static feature crossing or pooled behavior representations may neglect important sequential signals.

\paragraph{Token Mixing-based Recommendation Models.}
Instead of relying on heterogeneous handcrafted feature-crossing modules, emerging methods construct sequentialized unified features to exploit the scaling law of industrial-scale recommendation systems with large and non-stationary vocabularies \cite{zhai2024actions}. 
Derived from the MLP-driven alternatives to Vision Transformers \cite{tolstikhin2021mlp}, a line of recent work replaces quadratic self-attention with lightweight token mixing operators. These methods exchange information across tokens by mixing subspace vectors across tokens for global feature interactions \cite{chen2026rankup, ha2026unimixer}. For example, RankMixer \cite{zhu2025rankmixer} combines a multi-head token mixing module with per-token Feed-Forward Networks (FFNs) to model heterogeneous feature subspaces efficiently. TokenMixer-Large \cite{jiang2026tokenmixer} further improves this paradigm by redesigning the residual pathway with a mixing-and-reverting operation. 
Closely related to our design, MixFormer~\cite{huang2026mixformer} couples a Query Mixer over non-sequential features with cross-attention to behavior sequences. Its history pathway transforms each behavior independently using layer-specific FFNs. In contrast, CCFormer explicitly models interactions among neighboring behaviors through local joint token-channel mixing and hierarchically compresses the resulting contextualized representations. The resulting history hierarchy remains candidate-independent, allowing user and target queries to retrieve information at successive resolutions from fine-grained behaviors to broader historical context, while reducing subsequent computation. 

\paragraph{Attention-based Recommendation Models.}
Early sequential models tend to adopt recurrent neural networks, convolutional networks or attention mechanisms to encode historical behaviors \cite{kang2018self, sun2019bert4rec, zhai2024actions, yan2025scaling}. These methods significantly improve the expressiveness of user behavior modeling and have been widely explored in both retrieval and ranking scenarios. 
Recent studies further extend attention-based architectures to industrial-scale recommendation by jointly modeling behavior sequences, item features and user profiles within scalable frameworks \cite{zhang2026onetrans, guan2026make, lai2026unleashing, mtgr2025}. These works suggest that scaling up sequence length, model capacity and feature-interaction modules can consistently improve recommendation performance, especially when user interests are distributed across diverse historical behaviors and feature fields \cite{lin2025iterative, ding2026bending}.
Dense self-attention requires quadratic pairwise computation in sequence length, while its practical memory and latency costs depend on the implementation. Existing methods \cite{guan2026make, chai2025longer} make long histories tractable through query-based retrieval, selective attention, or token compression. CCFormer separates cross-field retrieval from local history contextualization and combines these operations with hierarchical compression across the block stack.

\section{Methods}

\begin{figure*}
    \centering
    \includegraphics[width=\textwidth]{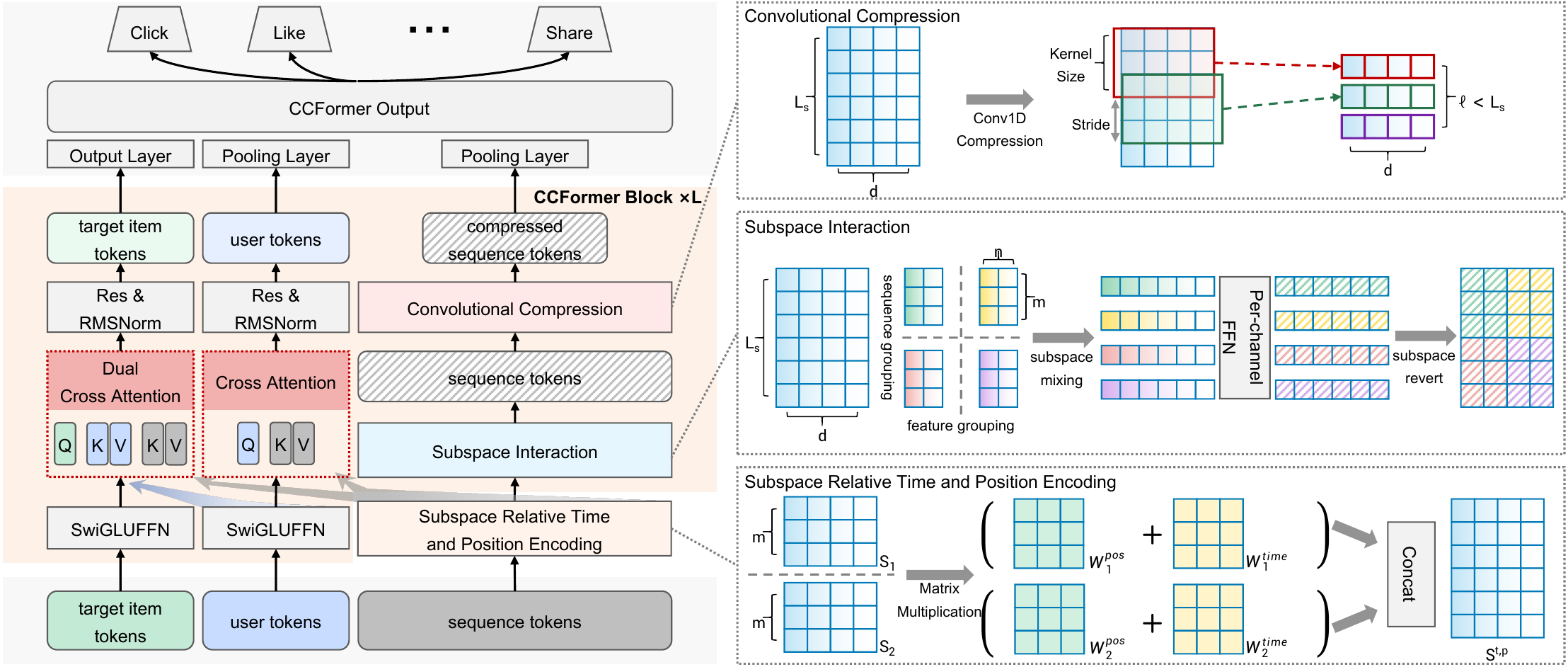}
    \caption{The overview of CCFormer.}
    \label{fig:2}
\end{figure*}

This section details the architecture of the proposed CCFormer. As depicted in Figure~\ref{fig:2}, CCFormer adopts a decoupled architectural paradigm. It explicitly partitions the input feature space into three distinct semantic fields: user profiles, historical behavior sequences, and target items. A fundamental limitation of models based on vanilla self-attention is the prohibitive quadratic computational cost of attending globally to all concatenated tokens. To overcome this bottleneck, CCFormer disentangles the feature interaction process into two orthogonal components: cross-field interaction and intra-sequence token interaction. This design enables the model to effectively distill target-aware user interests from extensive behavior histories while avoiding the quadratic cost of exhaustive self-attention over the full token set.

Formally, let $\mathbf{U}^{(0)}\in\mathbb{R}^{B\times L_u\times d}$, $\widetilde{\mathbf{S}}^{(0)}\in\mathbb{R}^{B\times L_s\times d}$, and $\mathbf{T}^{(0)}\in\mathbb{R}^{B\times L_t\times d}$ denote the initial tokenized representations for the user, behavior sequence, and target item fields, respectively. Here, $B$ represents the batch size, $d$ is the hidden dimensionality, and $L_u$, $L_s$, and $L_t$ denote the number of tokens in the respective fields. Specifically, heterogeneous user-side features are projected into a compact set of user tokens, whereas each historical behavior and target item is embedded as a fine-grained item-level token. Relative temporal-positional encoding is applied once to $\widetilde{\mathbf{S}}^{(0)}$, producing the initial behavior sequence token $\mathbf{S}^{(0)}$.

The core of CCFormer consists of $L$ stacked interaction blocks. Each block iteratively refines the representations of the three fields through the aforementioned decoupled interactions:
\begin{equation}
    (\mathbf{U}^{(\ell+1)},\mathbf{S}^{(\ell+1)},\mathbf{T}^{(\ell+1)})
    =\operatorname{CCFormerBlock}_{\ell}(\mathbf{U}^{(\ell)},\mathbf{S}^{(\ell)},\mathbf{T}^{(\ell)}),
\end{equation}
 After $L$ interaction blocks, the final normalized user and history representations are combined with each candidate's target representation and passed to task-specific prediction heads separately for that candidate, without aggregating across candidate targets.

 \subsection{Feature-Field Separated Cross Attention}

CTR prediction involves heterogeneous feature interactions, such as user-to-history preference retrieval, target-to-history relevance matching, and target-to-user compatibility modeling. Dense self-attention over concatenated fields allows interactions beyond these three retrieval relations and couples attention weights through shared normalization. CCFormer instead uses three directed cross-attention pathways to explicitly parameterize the selected relations.

Before cross-field interaction, we first apply lightweight token-wise feed-forward networks to the user and target fields:
\begin{equation}
    \bar{\mathbf{U}}^{(\ell)}=\operatorname{SwiGLUFFN}_{u}^{\mathrm{\ell}}(\mathbf{U}^{(\ell)}),\quad
    \bar{\mathbf{T}}^{(\ell)}=\operatorname{SwiGLUFFN}_{t}^{\mathrm{\ell}}(\mathbf{T}^{(\ell)}).
\end{equation}
The user field generates queries, keys, and values, while the behavior sequence only provides keys and values since it is used as the context to be retrieved:
\begin{equation}
    (\mathbf{Q}_{u},\mathbf{K}_{u},\mathbf{V}_{u})=\bar{\mathbf{U}}^{(\ell)}\mathbf{W}_{u}^{QKV},\quad
    (\mathbf{K}_{s},\mathbf{V}_{s})=\mathbf{S}^{(\ell)}\mathbf{W}_{s}^{KV}.
\end{equation}
Here, $\mathbf{W}_{u}^{QKV} \in \mathbb{R} ^ {d \times 3d}$ and $\mathbf{W}_{s}^{KV} \in \mathbb{R} ^ {d \times 2d}$.

The target field uses two query projections to attend to the behavior and user fields, respectively:
\begin{equation}
    \mathbf{Q}_{t\rightarrow s}=\bar{\mathbf{T}}^{(\ell)}\mathbf{W}_{t\rightarrow s}^{Q},\quad
    \mathbf{Q}_{t\rightarrow u}=\bar{\mathbf{T}}^{(\ell)}\mathbf{W}_{t\rightarrow u}^{Q}.
\end{equation}

The directed cross-attention is then computed as
\begin{equation}
\begin{aligned}
    \mathbf{O}_{u\rightarrow s}
    &= \operatorname{Attn}(\mathbf{Q}_{u},\mathbf{K}_{s},\mathbf{V}_{s};\mathbf{M}), \\
    \mathbf{O}_{t\rightarrow s}
    &= \operatorname{Attn}(\mathbf{Q}_{t\rightarrow s},\mathbf{K}_{s},\mathbf{V}_{s};\mathbf{M}), \\
    \mathbf{O}_{t\rightarrow u}
    &= \operatorname{Attn}(\mathbf{Q}_{t\rightarrow u},\mathbf{K}_{u},\mathbf{V}_{u}),
\end{aligned}
\end{equation}
where
\begin{equation}
    \operatorname{Attn}(\mathbf{Q},\mathbf{K},\mathbf{V};\mathbf{M})
    =
    \operatorname{softmax}
    \left(
    \frac{\mathbf{Q}\mathbf{K}^{\top}}{\sqrt{d_k}}+\mathbf{M}
    \right)\mathbf{V}.
\end{equation}
For each attention head, let $d_k = d/h$ denote the head dimension. The mask $\mathbf{M}$ is omitted when attention is applied without masking, and is used to mask padded behavior positions in sequence-related attention.

The user and target fields are updated by
\begin{equation}
    \Delta\mathbf{U}^{(\ell)}=\mathbf{O}_{u\rightarrow s}\mathbf{W}_{u}^{O},\quad
    \Delta\mathbf{T}^{(\ell)}=
    [\mathbf{O}_{t\rightarrow s};\mathbf{O}_{t\rightarrow u}]
    \mathbf{W}_{t}^{O}.
\end{equation}
Finally, we apply RMSNorm \cite{DBLP:conf/nips/ZhangS19a} to the cross-attention updates before adding them to the corresponding field-specific FFN outputs to obtain the user and target representations for the next block.
\begin{equation}
    \mathbf{U}^{(\ell+1)}=\operatorname{RMSNorm}( \Delta\mathbf{U}^{(\ell)}) + \bar{\mathbf{U}}^{(\ell)}
\end{equation}
\begin{equation}
\mathbf{T}^{(\ell+1)}=\operatorname{RMSNorm}( \Delta\mathbf{T}^{(\ell)}) + \bar{\mathbf{T}}^{(\ell)}
\end{equation}
Architecturally, this design introduces a relation-wise decomposition of cross-field interaction, explicitly modeling all three field pairs: user–history, target–history, and target–user. Each relation uses a dedicated query projection and a separately normalized attention map, while retaining shared context-side key/value representations where applicable. In particular, the target field evaluates user compatibility and historical relevance through distinct attention pathways, rather than a single attention distribution over concatenated user and history tokens. This prevents the two context fields from competing directly within the same softmax normalization pool, which is especially relevant when their token counts differ substantially. The relation-specific outputs are subsequently integrated through the field-update projections, separating relation estimation from evidence fusion. 

\subsection{Relative Temporal-Positional Encoding in Token Subspaces}

Before the first block, CCFormer encodes temporal proximity and relative order within local groups, aligning input contextualization with the first subspace mixer's interaction neighborhoods. Building on temporal-positional modeling~\cite{zhai2024actions,lai2026unleashing}, we partition $\widetilde{\mathbf{S}}^{(0)}$ into $m$-behavior groups and adapt the exponential-power kernel~\cite{yi2026fuxi}:
\begin{equation}\label{eq:4_2_1}
    \mathbf{W}^{\mathrm{time}}_{p,i,j} = \alpha \cdot \beta^{\lvert t_{p,i} - t_{p,j} \rvert^\gamma},
\end{equation}
where $t_{p,i}$ is the timestamp at position $i$ of group $p$, $\alpha,\gamma>0$ are learnable, and $\beta\in(0,1)$ controls decay. Smaller pairwise time gaps yield larger temporal weights. A learnable signed-offset table $\mathbf{W}^{\mathrm{pos\_origin}}\in\mathbb{R}^{2m-1}$ complements this symmetric proximity measure with relative order:
\begin{equation}\label{eq:4_2_3}
    \mathbf{W}^{\mathrm{pos}}_{p,i,j} = \mathbf{W}^{\mathrm{pos\_origin}}\left[ i-j+m-1 \right].
\end{equation}
Both matrices have shape $m\times m$. The positional table is shared across groups, whereas temporal weights vary with each user's observed behavior intervals. Writing $\widetilde{\mathbf{S}}^{(0)}_p$ for the $p$-th input group, we initialize the history state as
\begin{equation}\label{eq:4_2_5}
\begin{aligned}
    \mathbf{S}^{(0)}_p
    &=\left(\mathbf{W}^{\mathrm{time}}_p+\mathbf{W}^{\mathrm{pos}}_p\right)
    \widetilde{\mathbf{S}}^{(0)}_p,\\
    \mathbf{S}^{(0)}
    &=\operatorname{Concat}_{p=1}^{L_s/m}\left(\mathbf{S}^{(0)}_p\right).
\end{aligned}
\end{equation}

By restricting relation modeling to local groups, it requires only $O(L_s m)$ pairwise coefficients rather than a dense $O(L_s^2)$ relation matrix. This construction performs local temporal-positional aggregation rather than probability-normalized attention: the combined temporal and positional weights are not normalized, and the positional weights may be signed. Its local grouping aligns the granularity of input contextualization with that of subsequent token-channel interaction.

\subsection{Long-Sequence Subspace Token Mixing}

Cross-field attention retrieves historical evidence but does not directly mix behavior tokens within the history representation. CCFormer complements this retrieval with \emph{local joint token-channel mixing}, adapting subspace mixing~\cite{zhu2025rankmixer,jiang2026tokenmixer} to contiguous behavior groups. We reorganize $\mathbf{S}\in\mathbb{R}^{B\times L_s\times d}$ into:
\begin{equation}
    \mathbf{X}=\operatorname{Reshape}(\mathbf{S})\in\mathbb{R}^{B\times \frac{L_s}{m}\times \frac{d}{n}\times mn},
\end{equation}
where each $\mathbf{X}_{b,p,c}\in\mathbb{R}^{mn}$ packs $m$ adjacent behaviors and $n$ channels. This layout exposes cross-behavior, cross-channel combinations to a single nonlinear transformation, rather than applying a channel FFN to each behavior independently.

A channel-group-specific feed-forward network, denoted by $\operatorname{PFFN}_{c}$, performs gated mixing within each joint token-channel subspace:
\begin{equation}
    \operatorname{PFFN}_{c}(\mathbf{x})=
    \left(\phi(\mathbf{x}\mathbf{W}_{c}^{g})\odot (\mathbf{x}\mathbf{W}_{c}^{v})\right)\mathbf{W}_{c}^{o},
\end{equation}
where $\phi(\cdot)$ is a smooth activation and $\odot$ denotes element-wise multiplication. The projections $\mathbf{W}_{c}^{g}$, $\mathbf{W}_{c}^{v}$, and $\mathbf{W}_{c}^{o}$ jointly transform token and channel coordinates, enabling input-dependent modulation of local behavioral combinations:
\begin{equation}
    \mathbf{Z}_{b,p,c}=\operatorname{PFFN}_{c}(\mathbf{X}_{b,p,c}).
\end{equation}
Within each layer, parameters are distinct across channel groups but shared across sequence groups, allowing subspace specialization while reusing local interaction functions throughout the history. We restore the original sequence layout without reducing its token count:
\begin{equation}
    \mathbf{\hat{\mathbf{S}}}=\operatorname{Restore}(\mathbf{Z})\in\mathbb{R}^{B\times L_s\times d}.
\end{equation}

Crucially, the mixing width $mn$ is independent of $L_s$: longer histories increase the number of groups, not the size of each learned transformation. With fixed group sizes, feature dimensionality, and FFN width, mixing cost scales linearly with sequence length. The resulting interaction-enriched tokens support subsequent hierarchical compression, which broadens the scope of local mixing across depth.

\begin{figure}[!htbp]
    \includegraphics[width=\linewidth]{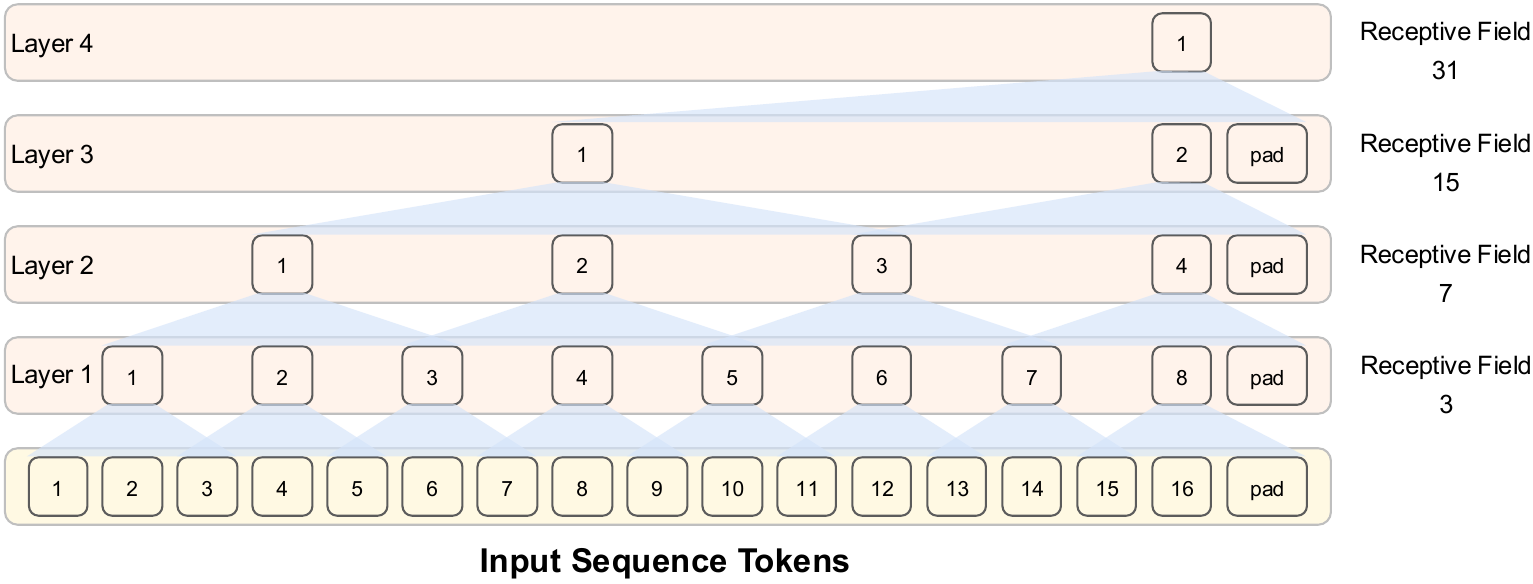}
    \caption{Change in sequence token receptive field with increasing model layers, Each layer in the figure denotes a compression stage.}\label{fig:3}
\end{figure}

\subsection{Hierarchical Long-Sequence Token Compression}

Long histories require a balance between retaining fine-grained behavioral evidence and limiting the cost of repeated interaction. CCFormer follows an \emph{interact-then-compress} hierarchy: each block performs cross-field retrieval and local subspace mixing at the current history resolution, while compression is applied at designated block boundaries. Let $\mathcal{C}$ denote the set of zero-based block indices followed by compression. The history update is
\begin{equation}
    \mathbf{S}^{(\ell+1)} =
    \begin{cases}
        \operatorname{Conv1D}_{k,s}\left(\hat{\mathbf{S}}^{(\ell)}\right), & \ell\in\mathcal{C},\\
        \hat{\mathbf{S}}^{(\ell)}, & \ell\notin\mathcal{C}.
    \end{cases}
\end{equation}
Here, $\hat{\mathbf{S}}^{(\ell)}$ is the history representation after subspace token mixing in block $\ell$, and $k$ and $s$ denote the convolution kernel size and stride. In the default eight-block configuration, $\mathcal{C}=\{1,3,5,7\}$, corresponding to compression after every two blocks.

The architectural value of this design lies in the coupling between \emph{interaction scope} and \emph{representation resolution}. Although each subspace mixer operates on only $m$ adjacent tokens, these tokens summarize increasingly broad spans of the original history after successive compression stages. Mixing at deeper layers therefore combines coarser behavioral units, allowing information from separate fine-resolution groups to interact at a broader scale. As illustrated in Figure~\ref{fig:3}, this composition expands receptive fields across depth without requiring dense pairwise attention over the original sequence. Compression thus changes the granularity of subsequent feature interaction, rather than merely reducing the token count.

Importantly, user and target queries read the evolving sequence hierarchy at successive blocks instead of accessing only a final compressed summary. Early layers can retrieve fine-grained behavioral evidence, while later layers query more contextualized representations with broader receptive fields. The history hierarchy remains candidate-independent, allowing different target queries to reuse the same compressed representations. Meanwhile, progressively shorter sequences reduce the work of subsequent subspace mixing and sequence-related cross-attention. Together, these properties make hierarchical compression a mechanism for coordinating multi-resolution feature interaction and computational allocation within the ranking backbone.

\subsection{Training and Deployment Optimization}

To support efficient training and deployment in industrial production environments, we introduce several practical optimizations for CCFormer. These optimizations improve computational efficiency and reduce resource overhead as detailed below.

\textbf{Mixed precision training.} We adopt mixed precision training (BF16 or FP16) for the main model computation. In practice, this strategy reduces memory overhead by more than 35\%, enabling larger batches or larger model parameters under the same training budget.

\textbf{Sparse parameters compression.} We store the embedding-table parameters of user and item ID features using INT8 symmetric linear quantization. We also apply a double-hashing strategy \cite{DBLP:conf/nips/SvenstrupHW17} to reduce the size of the ID-feature embedding tables. 

\textbf{Parallel prediction of candidate items.} Within the described CCFormer blocks, target tokens serve only as attention queries and do not attend to one another. This structure permits within-request reuse of user and history computations when their inputs are candidate-independent and the prediction heads operate separately on each candidate. During online serving, all candidate targets in a request are packed into the target field and forwarded only once. Since user field and sequence field modeling is shared across candidates, CCFormer produces multi-task scores for all targets in a single pass, avoiding repeated user sequence computation and improving serving throughput. In the reported deployment comparison with the pointwise DLRM baseline, the CCFormer serving configuration achieves 30\% higher peak QPS under the same inference resources. This is a system-level comparison and does not isolate the contribution of candidate packing from other architectural or deployment differences.

\section{Experiments}

\begin{table}[!ht]
    \centering
    \caption{Dataset Statistics.}\label{tab:dataset_statistic}
    \begin{tabular}{l|ccc}
        \toprule
        \textbf{Datasets} & \textbf{\#Users} & \textbf{\#Items} & \textbf{\#Raw interactions} \\
        \midrule
        Taobao   & 987,994 & 4,162,024 & 100,150,807   \\
        KuaiRec  & 7,176 & 10,728 & 12,530,806  \\
        Industry & >30M & >10M & >4B  \\
        \bottomrule
    \end{tabular}
\end{table}

\paragraph{Datasets.}
We conduct comparison experiments on two benchmarks (Taobao\footnote{\url{https://tianchi.aliyun.com/dataset/649}} and KuaiRec\footnote{\url{https://kuairec.com/}}) and an internal industry dataset. The dataset statistics are listed in Table~\ref{tab:dataset_statistic}. 
For the public datasets, we predict dataset-specific binary target labels from user interaction histories under the preprocessing protocol described in Appendix A. This protocol includes in-situ target replacement to construct synthetic negatives, so the public evaluations are not standard CTR prediction on natural exposure labels. The KuaiRec dataset is a short-video recommendation dataset, providing rich user-item interaction records. We use the provided big matrix data and treat an interaction as positive if its watch ratio is greater than 2. For both public datasets, the behavior sequence is truncated at
length 200 unless otherwise specified.
The industry dataset is collected from the online logs and real user feedback of a large-scale recommendation system. It contains one month of production data, comprising more than 4 billion samples, over 30 million users and 10 million recommended contents. For this industry dataset, the user behavior sequence length is set to 1000.

\begin{table*}[t]
    \centering
    \caption{Performance comparison on public datasets. $\Delta$AUC denotes the relative gain above chance (\%) with DIN as the reference, not an absolute percentage-point difference.}\label{tab:performance_public}
    \begin{tabular}{cc|ccc|cccc|c}
        \toprule
        \textbf{Datasets} & \textbf{Metric} & \textbf{DIN} & \textbf{DeepFM} & \textbf{MIMN} & \textbf{SASRec} & \textbf{HSTU} & \textbf{OneTrans} & \textbf{STCA} & \textbf{CCFormer} \\
        \midrule
        \multirow{2}{*}{Taobao}  & AUC (\%) & 88.33$\pm$0.22 & 89.06$\pm$0.42 & 91.79$\pm$0.32 & 88.52$\pm$0.65 & 91.40$\pm$0.25 & 91.35$\pm$0.35 & \second{92.81$\pm$0.38} & \first{93.67$\pm$0.32} \\
        & $\Delta$AUC (\%) & 0.00 & 1.90 & 9.03 & 0.50 & 8.01 & 7.88 & \second{11.69} & \first{13.93} \\
        \midrule
        \multirow{2}{*}{KuaiRec} & AUC (\%) & 80.22$\pm$0.58 & 80.14$\pm$0.51 & 81.79$\pm$0.81 & 79.83$\pm$0.70 & 82.62$\pm$0.39 & \second{82.76$\pm$0.32} & 82.18$\pm$0.16 & \first{83.35$\pm$0.29} \\
        & $\Delta$AUC (\%) & 0.00 & -0.26 & 5.20 & -1.29 & 7.94 & \second{8.41} & 6.49 & \first{10.36} \\
        \bottomrule
    \end{tabular}
\end{table*}

\begin{table*}[t]
    \centering
    \caption{Performance comparison on the industrial dataset. $\Delta$AUC and $\Delta$GAUC denote relative gains above chance (\%) with HSTU as the reference, not absolute percentage-point differences.}\label{tab:performance_industry}
    \newcommand{\industrytablecontent}{%
        \toprule
        \textbf{Model arch type} & \textbf{Models} & \textbf{\makecell{AUC \\ (\%)}} & \textbf{\makecell{$\Delta$AUC \\ (\%)}} & \textbf{\makecell{GAUC \\ (\%)}} & \textbf{\makecell{$\Delta$GAUC \\ (\%)}} & \textbf{\makecell{Dense \\ Params(M)}} & \textbf{GFLOPs/Sample} \\
        \midrule
        Unified-Block Model & HSTU     & 77.66 & 0.00 & 70.86 & 0.00 & 19.84 & 36.89 \\
        Unified-Block Model & OneTrans & 77.69 & 0.11 & 70.90 & 0.19 & 20.21 & 35.58 \\
        Unified-Block Model & STCA     & 77.73 & 0.25 & 70.95 & 0.43 & 20.17 & 26.44 \\
        \midrule
        Two-Stage Model & HSTU+RankMixer  & 77.74 & 0.29 & 70.95 & 0.43 & 20.32 & 40.60 \\
        Two-Stage Model & HSTU+TokenMixer & \second{77.76} & \second{0.36} & \second{71.03} & \second{0.81} & 20.35 & 41.45 \\
        \midrule
        Unified-Block Model & CCFormer & \first{77.94} & \first{1.01} & \first{71.36} & \first{2.40} & 20.07 & 16.64 \\
        \bottomrule
    }
    \setbox0=\hbox{\begin{tabular}{c|c|cc|cc|cc}\industrytablecontent\end{tabular}}
    \setlength{\tabcolsep}{\dimexpr\tabcolsep+(\textwidth-\wd0)/16\relax}
    \begin{tabular}{c|c|cc|cc|cc}
        \industrytablecontent
    \end{tabular}
\end{table*}

\paragraph{Baselines.}
For the public datasets, we compare CCFormer with three representative baselines under the
conventional DLRM paradigm (DIN \cite{zhou2018deep}, DeepFM \cite{guo2017deepfm} and MIMN \cite{pi2019practice}) and four attention-based sequential recommendation models (HSTU \cite{zhai2024actions}, OneTrans \cite{zhang2026onetrans}, SASRec \cite{kang2018self} and STCA \cite{guan2026make}). We repeat each experiment 5 times to report AUC and the corresponding relative improvement. 
On the industrial dataset, we further compare CCFormer with strong production-oriented baselines (HSTU, OneTrans and STCA) using both AUC and Group AUC (GAUC) as evaluation metrics. 
Following \cite{xu2020deep}, $\Delta\mathrm{AUC}$ and $\Delta\mathrm{GAUC}$ denote relative gains above the chance level of 0.5, rather than absolute percentage-point differences or ordinary relative changes. For metric values expressed on the $[0,1]$ scale, we compute $\mathrm{RelaImpr}=\left(\frac{\mathrm{Metric}(\mathrm{measured\ model})-0.5}{\mathrm{Metric}(\mathrm{base\ model})-0.5}-1\right)\times 100\%$. Equivalently, when using the percentage-scale values shown in the tables, the chance-level reference is 50 rather than 0.5.

\paragraph{Implementation Details.}
Unless otherwise stated, for all industrial experiments, we set the feature-token dimensionality to $d=256$  and stack 8 CCFormer blocks. The subspace group size and channel group size are set to $m=8$ and $n=16$, respectively. The convolutional compression module uses a kernel size of $k=3$ and stride of $s=2$, and sequence compression is applied once every two blocks. All models are optimized using Adam with a learning rate of $1\times10^{-4}$ and a batch size of $4096$. All experiments are conducted using the same hardware and software configuration on a cluster of 16 NVIDIA H20 GPUs.

\begin{figure}[t]
\centering
  \includegraphics[width=0.8\linewidth]{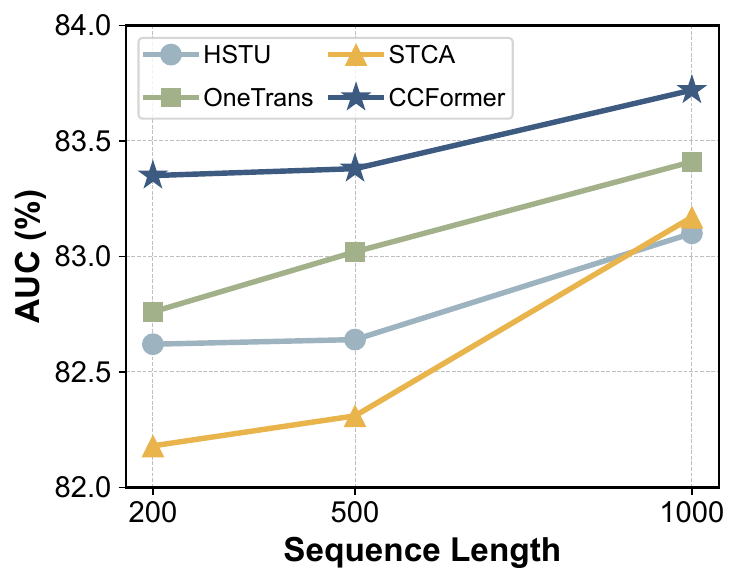}
\caption{AUC comparison of four sequential recommendation models under increasing sequence lengths on KuaiRec.}
\label{fig:4}
\end{figure}

\begin{figure*}[t]
\centering
    \includegraphics[width=\linewidth]{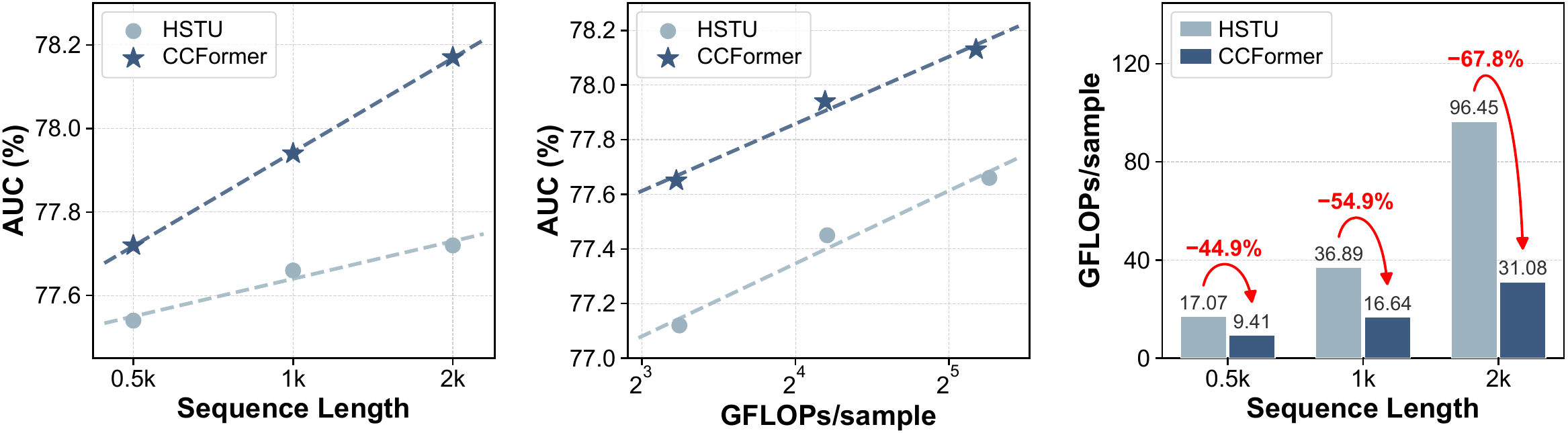}
\caption{Scaling analysis of CCFormer and HSTU. Left: AUC at matched sequence lengths. Middle: AUC versus GFLOPs per sample across the evaluated configurations, with sequence length held fixed. Right: GFLOPs per sample at different sequence lengths under comparable parameter counts.}
\label{fig:5}
\end{figure*}

\subsection{Overall Performance}
Tables~\ref{tab:performance_public} and~\ref{tab:performance_industry} show that \textbf{CCFormer achieves the best predictive performance among the evaluated models on both public and industrial datasets.} On Taobao and KuaiRec, it achieves AUCs of 93.67\% and 83.35\%, exceeding the strongest respective baselines, STCA and OneTrans, by 0.86 and 0.59 percentage points. These results show the highest reported mean AUC among the evaluated models on both public datasets under the adopted preprocessing protocols. We further compare four advanced sequential recommendation models under different sequence lengths (200, 500 and 1000), as shown in Figure~\ref{fig:4}. CCFormer achieves higher reported AUC than HSTU, OneTrans, and STCA at each of the three evaluated sequence lengths on KuaiRec.

The industrial comparison covers two architectural categories: \emph{unified-block models} (HSTU, OneTrans, STCA, and CCFormer) and \emph{two-stage models} (HSTU+RankMixer and HSTU+TokenMixer). CCFormer achieves 77.94\% AUC and 71.36\% GAUC, improving over HSTU by 0.28 and 0.50 percentage points, respectively, or 1.01\% and 2.40\% under the relative-improvement definition. Both two-stage baselines improve upon HSTU, while CCFormer exceeds the strongest baseline, HSTU+TokenMixer, by 0.18 and 0.33 percentage points in AUC and GAUC, respectively.

\textbf{CCFormer also achieves the lowest GFLOPs per sample among the compared models.} With 20.07M parameters, compared with 20.35M for HSTU+TokenMixer, CCFormer requires 16.64 GFLOPs per sample, versus 41.45 for HSTU+TokenMixer and 36.89 for HSTU, corresponding to reductions of 59.9\% and 54.9\%, respectively. Even relative to STCA, the baseline with the lowest GFLOPs, CCFormer reduces computation by 37.1\% while improving AUC and GAUC by 0.21 and 0.41 percentage points. Overall, CCFormer offers a favorable effectiveness-efficiency trade-off relative to both architectural categories under the evaluated configurations.

\subsection{Scaling Analysis of CCFormer}
In this section, we examine CCFormer's empirical accuracy-compute trends over the evaluated sequence lengths and model configurations. We focus on two factors, sequence length and model GFLOPs per sample, which are crucial for sequential recommendation models. All experiments are conducted under the same hardware and software stack. In addition to performance metrics, we report the corresponding GFLOPs of each model.

\paragraph{Sequence Length Scaling.} 
We first investigate whether CCFormer can effectively scale to longer user behavior sequences on the industrial dataset, matching the input sequence length between the two models at each setting. As shown in Figure~\ref{fig:5}, \textbf{CCFormer consistently outperforms HSTU under different sequence lengths.} 
As the sequence length increases, CCFormer's AUC improves steadily, and its performance gap over HSTU widens. This suggests that CCFormer benefits more from longer behavior histories. \textbf{CCFormer with only 0.5k behavior tokens already achieves AUC comparable to HSTU with 2k behavior tokens.} These trends are consistent with the design of CCFormer. Long-sequence subspace token mixing captures expressive patterns from full behavior histories. Sequence compression preserves a large receptive field at substantially lower cost.

\paragraph{GFLOPs Scaling.}
We next compare AUC and GFLOPs per sample across the evaluated model configurations while holding the input sequence length fixed.
Figure~\ref{fig:5} summarizes the observed accuracy-compute trade-offs across these configurations. 
We then examine how computational cost scales with sequence length using model configurations with comparable dense parameter counts. As the sequence length grows from 500 to 2000, CCFormer's computational cost increases from 9.41 to 31.08 GFLOPs per sample, compared with 17.07 to 96.45 for HSTU. At matched sequence lengths, the computational savings widen from 44.9\% at 500 tokens to 54.9\% at 1000 tokens and 67.8\% at 2000 tokens. These results show a favorable observed accuracy-compute trade-off for CCFormer across the evaluated configurations and an increasing reduction in GFLOPs per sample at longer sequence lengths under comparable dense parameter counts.

\begin{figure}[t]
\centering
  \includegraphics[width=\linewidth]{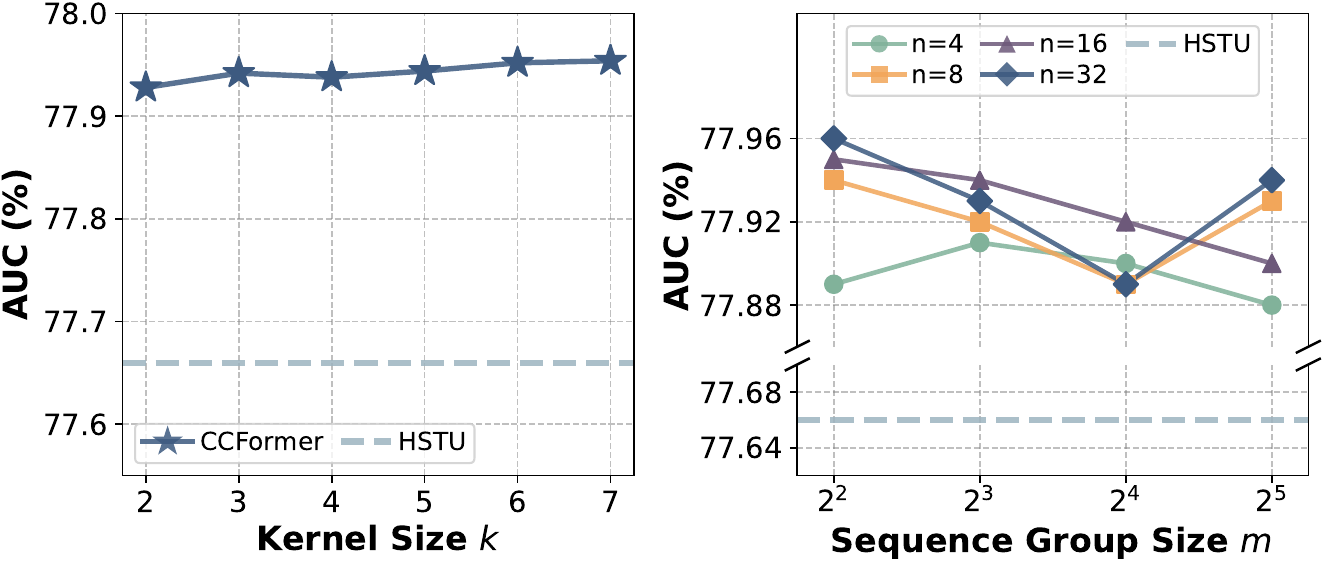}
\caption{AUC sensitivity to the convolution kernel size and the sequence-group and channel-group sizes in CCFormer under the evaluated configurations.}
\label{fig:6}
\end{figure}

\begin{table*}[tbp]
    \centering
    \caption{Ablation results on the industrial dataset. HSTU is the reference for the reported training-speed ratios and relative gains above chance in AUC and GAUC.}\label{tab:ablation_study}
    \begin{tabular*}{\textwidth}{@{\extracolsep{\fill}}c|c|c|cc|cc}
        \toprule
        \textbf{Models} & \textbf{Note} & \textbf{\makecell{Training \\ throughput ratio}} & \textbf{\makecell{AUC \\ (\%)}} & \textbf{\makecell{$\Delta$AUC \\ (\%)}} & \textbf{\makecell{GAUC \\ (\%)}} & \textbf{\makecell{$\Delta$GAUC \\ (\%)}} \\
        \midrule
        HSTU & - & 1.00 & 77.66 & 0.00 & 70.86 & 0.00 \\
        \midrule
        \multirow{9}{*}{CCFormer} & - & 2.21 & \second{77.94} & \second{1.01} & \first{71.36} & \first{2.40} \\
        & Remove relative temporal-positional encoding & 2.35 & 77.85 & 0.69 & 71.13 & 1.29 \\
        & Remove feature-field separated cross attention & 2.59 & 77.38 & $-$1.01 & 70.56 & $-$1.44 \\
        & Feature-field separated cross attention $\rightarrow$ Unified self-attention & 1.96 & 77.82 & 0.58 & \second{71.25} & \second{1.87} \\
        & Long-sequence token mixing $\rightarrow$ Self-attention & 1.41 & \first{77.96} & \first{1.08} & 71.21 & 1.68 \\
        & Long-sequence token mixing $\rightarrow$ Compressed sparse attention & 1.74 & 77.80 & 0.51 & 71.20 & 1.63 \\
        & Remove long-sequence token mixing & 2.56 & 77.73 & 0.25 & 71.02 & 0.77 \\
        & Remove sequence compression & 1.29 & 77.92 & 0.94 & \second{71.25} & \second{1.87} \\
        & Sequence compression $\rightarrow$ Mean pooling & 2.24 & 77.76 & 0.36 & 71.08 & 1.05 \\
        \bottomrule
    \end{tabular*}
\end{table*}

\begin{table*}[!htbp]
    \centering
    \small
    \setlength{\tabcolsep}{2pt}
    \renewcommand{\arraystretch}{0.95}
    \setlength{\aboverulesep}{1pt}
    \setlength{\belowrulesep}{1pt}
    \caption{Online A/B gains of CCFormer.}\label{tab:ab_test}
    \begin{tabular*}{\textwidth}{@{\extracolsep{0pt}}l|c|c|c|l@{\extracolsep{\fill}\hspace{2\tabcolsep}}lll@{}}
        \toprule
        \textbf{Scenario} & \textbf{Baseline} & \textbf{$\Delta$ Seq length} & \textbf{$\Delta$ QPS} & \multicolumn{4}{c@{}}{\textbf{Online metric relative lifts}} \\
        \midrule
        \multirow{2}{*}{Scenario 1} & \multirow{2}{*}{DLRM} & \multirow{2}{*}{+500\%} & \multirow{2}{*}{+30\%} & CTR: 3.57\% & UCTR: 1.93\% & 3-day Activeness: 1.93\% & Video View: 3.47\% \\
        & & & & Unique Viewer: 2.61\% & Page View: 3.86\% & Watch Time: 1.29\% & Ad revenue: 1.64\% \\
        \midrule
        Scenario 2 & HSTU+RankMixer & 0\% & +34.5\% & \multicolumn{2}{l}{Ad View: 1.43\%} & \multicolumn{2}{l@{}}{Ad revenue: 1.71\%} \\
        \midrule
        Scenario 3 & HSTU+RankMixer & +65.1\% & +11.1\% & CTR: 2.02\% & UCTR: 1.67\% & New User Page View: 4.01\% & Ad revenue: 1.29\% \\
        \midrule
        Scenario 4 & HSTU & 0\% & +28.3\% & Page View: 0.45\% & Watch Time: 1.26\% & Share: 3.49\% & CTR: 1.01\% \\
        \bottomrule
    \end{tabular*}
\end{table*}

\subsection{Effect of Long-Sequence Modeling Hyperparameters}
To assess the robustness of CCFormer, we investigate two types of hyperparameters related to long-sequence modeling. Specifically, the kernel size $k$ in the sequence compression module controls the receptive field over neighboring behavior tokens. The group size is used in long-sequence subspace token mixing, where the sequence group size $m$ partitions behavior tokens along the sequence dimension and the channel group size $n$ partitions token embeddings along the feature dimension.

\paragraph{Kernel Size.} As shown in Figure~\ref{fig:6}, increasing the kernel size slightly improves AUC, while the overall performance remains stable. When the kernel size increases from 2 to 7, AUC changes from 77.92\% to 77.95\%. Moreover, CCFormer consistently outperforms HSTU by a clear margin under different kernel sizes. 
These results indicate that CCFormer maintains stable predictive performance across the evaluated kernel sizes. In the industrial experiments, we set the stride $s$ to 2 as a conservative choice that substantially reduces computation while retaining finer-grained behavioral resolution; the effect of the stride is further analyzed in Appendix~\ref{apdx:sensitivity}.

\paragraph{Group Size.} We then study the key hyperparameters, the sequence group size $m$ and channel group size $n$, in the long-sequence subspace token mixing module. As shown in Figure~\ref{fig:6}, CCFormer remains stable under different combinations of $m$ and $n$ and all configurations consistently outperform the HSTU baseline. Although smaller group sizes can slightly improve AUC in some cases, the overall variation is limited. These results show limited AUC variation across the evaluated sequence-group and channel-group configurations.

\subsection{Ablation Study}
Table~\ref{tab:ablation_study} reports eight ablations on the industrial dataset under the same training protocol. Here and in Appendix~\ref{apdx:sensitivity}, the training-throughput ratio is defined as the evaluated model's training throughput in samples per second divided by that of HSTU, with HSTU normalized to 1.00.
\paragraph{Cross-field interaction.}
Removing feature-field-separated cross-attention yields the highest training-throughput ratio relative to HSTU among the tested variants ($2.59\times$), but also the largest reduction in AUC/GAUC, to 77.38\%/70.56\%, below HSTU. This result indicates that cross-field retrieval is important in the evaluated architecture. Replacing the module with unified self-attention yields lower reported AUC/GAUC (77.82\%/71.25\%) and a lower training-throughput ratio ($1.96\times$) than the full model.
\paragraph{Sequence interaction.}
Removing relative temporal-positional encoding lowers the reported AUC/GAUC to 77.85\%/71.13\%. Removing subspace token mixing increases the training-throughput ratio relative to HSTU from $2.21\times$ to $2.56\times$, while lowering AUC/GAUC from 77.94\%/71.36\% to 77.73\%/71.02\%. Replacing token mixing with full self-attention yields a slightly higher reported AUC (77.96\%) but lower GAUC (71.21\%) and a lower training-throughput ratio ($1.41\times$); replacing it with CSA~\cite{xu2026deepseek} yields lower AUC/GAUC (77.80\%/71.20\%) and a ratio of $1.74\times$.
\paragraph{Sequence compression.}
Removing sequence compression reduces the training-throughput ratio relative to HSTU from $2.21\times$ to $1.29\times$, while AUC/GAUC change from 77.94\%/71.36\% to 77.92\%/71.25\%. These results support a substantial throughput benefit from compression without an observed accuracy penalty in this comparison. Replacing convolutional compression with mean pooling yields AUC/GAUC of 77.76\%/71.08\% and a training-throughput ratio of $2.24\times$.

\subsection{Online A/B Results}
We evaluate CCFormer across four production scenarios against DLRM (Scenario 1), HSTU+RankMixer (Scenarios 2 and 3), and HSTU (Scenario 4). Each experimental group receives over one million exposed users daily during two-week A/B tests. Statistical inference is performed at the user level, consistent with the randomization unit.

Table~\ref{tab:ab_test} shows consistent business gains across all four scenarios. In video recommendation (Scenario 1), CTR, Page View, and advertising revenue improve by 3.57\%, 3.86\%, and 1.64\%, respectively. Advertisement ranking (Scenario 2) achieves 1.43\% higher Ad View and 1.71\% higher revenue. Scenario 3 improves CTR by 2.02\%, New User Page View by 4.01\%, and revenue by 1.29\%. Scenario 4 improves Share, Watch Time, and CTR by 3.49\%, 1.26\%, and 1.01\%, respectively. For each scenario, QPS is compared between CCFormer and its corresponding online baseline under identical serving-resource allocations and the same upper bound on mean serving response time. Beyond business gains, Scenarios 1 and 3 increase sequence lengths by 500\% and 65.1\% while improving QPS by 30\% and 11.1\%, respectively. With unchanged sequence lengths, Scenarios 2 and 4 achieve QPS gains of 34.5\% and 28.3\%, respectively. All changes are relative to the corresponding online baselines. Each scenario undergoes staged evaluation with 5-10\% traffic per group, expanded testing at approximately 20\% per group, and a reverse-traffic experiment after full rollout. Deployment is finalized only when key metrics show sustained, statistically significant improvements throughout this process. Reported business-metric gains are statistically significant (two-sample $t$-tests, $p<0.05$). CCFormer is now fully deployed across all four scenarios, serving the entire production traffic.

\section{Conclusion}

We presented CCFormer, a structured candidate-ranking architecture that separates cross-field retrieval from local history modeling and combines them with hierarchical sequence compression. On the industrial dataset, CCFormer exceeds HSTU+TokenMixer by 0.18 and 0.33 percentage points in AUC and GAUC, respectively, while using 59.9\% fewer reported GFLOPs per sample under the evaluated configurations. Across the tested sequence lengths, it exhibits a favorable empirical accuracy-compute trade-off. CCFormer has been fully deployed in four production scenarios, where the deployed systems show business-metric gains and serving-QPS increases of 11.1\%-34.5\% relative to their respective baselines. These online results describe system-level outcomes rather than isolated effects of individual modules. Future work will investigate adaptive layer-wise compression and evaluate its applicability to longer behavior histories.

\bibliographystyle{ACM-Reference-Format}
\bibliography{reference}

\appendix

\section{Experimental Settings}

For the industrial dataset, we use real user samples collected over 31 consecutive days, where the first 30 days are used for training and the final day for testing. For both public datasets, we sort each user's interactions chronologically and construct samples consisting of a target interaction and its preceding behavior history. Following the previous study \cite{pi2019practice}, we use the $0.7$ quantile of the target timestamps as a global cutoff. Samples with target timestamps after the cutoff form the test set and the remaining samples form the training set. We further hold out a fraction of $0.1$ of this training set for validation and hyperparameter selection.
For Taobao, we construct one sample using the last interaction as the target and all preceding interactions as the behavior history. Besides, we perform in-situ target replacement with probability 0.5. We replace the target item with a uniformly sampled item from the dataset's item set and assign the resulting example a negative label \cite{pi2019practice}. All item-side features are updated consistently with the replacement item.
For KuaiRec \cite{gao2022kuairec}, we use the provided big matrix and exclude users with fewer than two interactions. We sort each user's interactions chronologically and construct one sample for every target interaction with a preceding history, allowing multiple samples per user. Each target interaction is labeled positive if its watch ratio is greater than 2, and negative otherwise. We retain the original negative samples derived from observed viewing feedback.
In addition, we adopt Tree-structured Parzen Estimator (TPE) \cite{bergstra2011tpe} to search for the optimal baseline configurations on the public datasets.

\section{Additional Sensitivity Analysis}\label{apdx:sensitivity}

We further investigate the influence of hierarchical sequence compression on predictive performance and training efficiency for the industrial dataset. We vary the Conv1D stride $s\in\{1,2,3,4\}$ while fixing the kernel size at $k=3$ and keeping all other settings at default values. Larger strides reduce the sequence length more aggressively, allowing us to examine the trade-off between computational savings and the retention of behavioral information. Note that the setting $s=1$ still retains the Conv1D operation, which performs learnable aggregation over neighboring tokens without reducing the sequence length. We report AUC and training-throughput ratios relative to HSTU under the same hardware and software configuration.

As shown in Table~\ref{tab:stride}, increasing the stride from 1 to 3 raises training speedup from $1.25\times$ to $2.46\times$. Meanwhile, reported AUC remains close across strides 1–3 as training throughput increases. At stride 4, AUC decreases, indicating a less favorable accuracy–throughput trade-off under the tested configuration. In a nutshell, setting the stride to either 2 or 3 offers a favorable balance between predictive performance and training efficiency.

\begin{table}[htbp]
  \centering
  \caption{Effect of stride on AUC and training-throughput ratio relative to HSTU.}
  \label{tab:stride}
  \begin{tabular}{lccc}
    \toprule
    Model & Stride & AUC (\%) & Throughput ratio ($\times$) \\
    \midrule
    HSTU     & - & 77.66 & $1.00\times$ \\
    \midrule
    \multirow{4}{*}{CCFormer} & 1  & 77.91 & $1.25\times$ \\
     & 2  & 77.94 & $2.21\times$ \\
     & 3  & 77.95 & $2.46\times$ \\
     & 4  & 77.85 & $2.61\times$ \\
    \bottomrule
  \end{tabular}
\end{table}

\section{Online Serving Configurations}\label{apdx:online_serving}

\begin{table}[]
  \centering
  \footnotesize
  \setlength{\tabcolsep}{2pt}
  \renewcommand{\arraystretch}{1.1}
  \caption{Online serving configurations and QPS gains of CCFormer.}
  \label{tab:online_serving_configurations}
  \begin{tabular*}{\columnwidth}{@{\extracolsep{\fill}}lcccc@{}}
    \toprule
    \textbf{Scenario} & \textbf{Baseline} & \makecell{\textbf{$\Delta$ Seq} \\ \textbf{length}} & \textbf{$\Delta$ QPS} & \makecell{\textbf{Baseline uses} \\ \textbf{parallel prediction} \\ \textbf{of candidate items?}} \\
    \midrule
    Scenario 1 & DLRM & +500\% & +30\% & NO \\
    \midrule
    Scenario 2 & \makecell{HSTU+ \\ RankMixer} & 0\% & +34.5\% & YES \\
    \midrule
    Scenario 3 & \makecell{HSTU+ \\ RankMixer} & +65.1\% & +11.1\% & YES \\
    \midrule
    Scenario 4 & HSTU & 0\% & +28.3\% & YES \\
    \bottomrule
  \end{tabular*}
\end{table}

Table~\ref{tab:online_serving_configurations} contextualizes the QPS gains in Table~\ref{tab:ab_test} by reporting whether each baseline already uses parallel prediction of candidate items. Within each scenario, QPS is compared under identical serving-resource allocations and the same upper bound on mean serving response time. For recommendation scenarios that already use parallel candidate prediction during online serving, CCFormer and the corresponding baseline process the same number of candidate items per inference call at the ranking stage. This number varies across production applications, ranging from approximately 500 to 2,000.

\paragraph{Comparison with a pointwise baseline.}
In Scenario 1, the DLRM baseline does not use parallel candidate prediction. CCFormer increases sequence length by 500\% while improving QPS by 30\%. This result demonstrates a favorable system-level trade-off when moving from pointwise serving to CCFormer's shared-computation design. However, candidate parallelism, model architecture, and sequence length all differ in this comparison; the 30\% QPS gain cannot be attributed to parallel prediction alone.

\paragraph{Comparisons with parallel baselines.}
The baselines in Scenarios 2-4 already use parallel candidate prediction. At unchanged sequence lengths, CCFormer improves QPS by 34.5\% over HSTU+RankMixer in Scenario 2 and by 28.3\% over HSTU in Scenario 4. In Scenario 3, it supports a 65.1\% longer sequence while still improving QPS by 11.1\% over HSTU+RankMixer. These results show that CCFormer's serving advantage persists against baselines that already support candidate parallelism, rather than arising solely from enabling this capability against a pointwise baseline.


\end{document}